\documentclass[twocolumn,amsmath,amssymb,prl,superscriptaddress,showpacs]{revtex4}
\usepackage{graphicx}
\usepackage{appendix}
\usepackage{bm}
\usepackage{braket}

\begin{document}

\title{Absence of  helical surface states in bulk semimetals with broken inversion symmetry}

\author{Carmine Ortix} 
\affiliation{Institute for Theoretical Solid State Physics, IFW Dresden, D01171 Dresden, Germany}
\author{J\"orn W.F. Venderbos} 
\affiliation{Institute for Theoretical Solid State Physics, IFW Dresden, D01171 Dresden, Germany}
\author{Roland Hayn}
\affiliation{Aix-Marseille Univ., CNRS, IM2NP-UMR 7334, 13397 Marseille Cedex 20, France} 
\author{Jeroen van den Brink} 
\affiliation{Institute for Theoretical Solid State Physics, IFW Dresden, D01171 Dresden, Germany}

\date{\today}

\begin{abstract}
Whereas the concept of topological band-structures was developed originally for insulators with a bulk bandgap, it has become increasingly clear that the prime consequences of a non-trivial topology -- spin-momentum locking of surface states --  can also be encountered in gapless systems. Concentrating on the paradigmatic example of mercury chalcogenides HgX (X = Te, Se, S), we show that the existence of  {\it helical semimetals}, i.e. semimetals with topological surface states, critically depends on the presence of  crystal inversion symmetry. An infinitesimally small broken inversion symmetry (BIS) renders the helical semimetallic state unstable. The BIS is also very important in the fully gapped regime, renormalizing the surface Dirac cones in an anisotropic manner. As a consequence the handedness of the Dirac cones can be flipped by a biaxial stress field. 
 \end{abstract}

\maketitle

\date{\today}

{\it Introduction -- }  The discovery of two- and three-dimensional (3D) topological insulators (TIs) \cite{kan05,kan05b,ber06,wu06,kon07,fu07,fu07b,moo07,zha09,hsi08,xia09,che09,hsi09,has10,ras13} has brought to light a new state of quantum matter. This has had a tremendous impact in the field of fundamental condensed matter physics as well as for potential applications in spintronics and quantum computation \cite{akh09}. The TIs are insulating in the bulk but have topologically protected surface states \cite{kan05b,wu06,has10} and the topology dictates that the metallic surface states are spin-momentum locked:  surface electrons with opposite spin counter-propagate at the sample boundaries \cite{ber06,wu06,fu07b,zha09}. 

Materials with a TI band structure such as 
Sb \cite{hsi09}, Bi$_2$Se$_3$ \cite{eto10}  and Bi$_{14}$Rh$_3$I$_9$ \cite{ras13}
often show the presence of a finite bulk carrier density. In such materials, the bulk Fermi surface {\it does not} simply swallow up the topological surface states.  They survive and coexist with a bulk Fermi surface \cite{hsi09}, leading to the notion of a {\it helical metal}. The coexistence of a bulk Fermi surface and topological surface states can be understood as a doped TI being made out of a bulk TI with a non-topological metallic band inside the gap. The hybridization between the topological surface states and the additional metallic band pushes the topological surface states away from overlapping with the bulk states in energy and momentum. This preserves them in a slightly modified form at those points in the Brillouin zone (BZ) where the surface and the bulk bands do not cross \cite{ber10}. Also in bulk semimetals with a topological non-trivial band ordering surface Dirac-like states are expected to coexist with metallic states \cite{bru11,chu11}, suggesting the analogous presence of a helical semimetallic state. 

Using the paradigmatic example of the series of cubic mercury chalcogenides HgX (X = Te, Se, S), we show however that the existence of a helical semimetallic state critically relies on the presence of crystal inversion symmetry even in the absence of disorder.  An infinitesimally small broken inversion symmetry (BIS) is detrimental for the topological surface states of a helical semimetal, {\it independent of} any overlap in energy and momentum of bulk and topological surface states. 
We show furthermore that in the fully gapped TI regime, a BIS does not endanger the existence of topological surface states. In this case the BIS rather renormalizes the Fermi velocity of the surface Dirac fermions in an anisotropic manner, similarly to the effect envisioned in anisotropic graphene superlattices \cite{par08np,par08nl}. This in principle allows an externally applied biaxial stress field to flip the surface state chirality in a material with BIS. 

\begin{figure}
 \includegraphics[width= .9 \columnwidth]{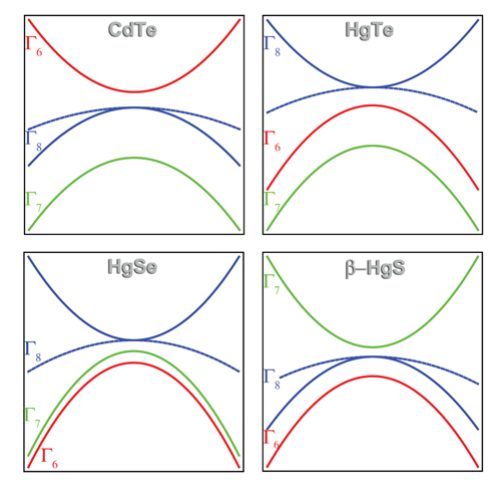}
 \caption{Schematic band-structure close to the Brillouin zone centre of the cubic mercury chalcogenides inverted semiconductors HgX as compared to the topologically trivial CdTe semiconductor.}
 \label{fig1}
 \end{figure} 

{\it HgX compounds } -- Pristine HgTe is a semimetal which is charge neutral when the Fermi energy is at the touching point between the light-hole (LH) and the heavy-hole (HH) $\Gamma_8$ bands at the BZ center \cite{dai08,chu11}. 
The topological nature of the electronic states in this material cannot be inferred from these $p_{3/2}$ atomic levels  but rather follows from 
the inverted band ordering at the zone center of  the LH $\Gamma_8$ band which is particle-like and the $\Gamma_6$ $s$-band which is hole-like. In normal semiconductors, such as CdTe [see Fig.1], $\Gamma_6$ forms the conduction band and $\Gamma_8$ is one of the valence bands. The consequence of this band inversion can be understood from the simple 
criterion
derived by Fu and Kane \cite{fu07} to distinguish normal and non-trivial topological classes.  
This criterion, which relies on the presence of inversion symmetry, establishes a material to be in a topologically non-trivial class if two bands of opposite parity have level-crossed with respect to the normal band ordering. The zinc-blende crystal structures of HgTe lacks inversion symmetry, but it is 
normally considered that the BIS
acts as a small perturbation and, by invoking the principle of adiabatic continuity, does not hinder the topological nature of the level crossing. 
As the HH bands do not participate in the topological level crossing, it can be assumed that they act as inserted "parasitic" bulk bands closing the full band gap and preventing the system to be a strong 3D TI. 
It is expected  \cite{bru11,chu11}  that the existence of topological surface states resulting from the LH-$\Gamma_6$ TI bulk is not undermined by the presence of the HH bulk bands suggesting HgTe to be a helical semimetal. 

Similar arguments apply to HgSe which has the same band ordering as HgTe but with the difference that the spin-orbit (SO) split-off $\Gamma_7$ bands are above the $\Gamma_6$ bands \cite{sva11} -- the SO splitting $\Delta_0=E(\Gamma_8) - E(\Gamma_7)$ is smaller than the gap $-E_0= E(\Gamma_8) - E(\Gamma_6)$ [c.f. Fig.~\ref{fig1}]. 
In this case the SO split-off bands, the LH and the $\Gamma_6$ bands realise a bulk TI with the HH bands playing as in HgTe the role of parasitic bands which close the full band gap. 
Yet another material of the  same family -- metacinnabar -- was proposed to be in a topologically non-trivial class: 
a recent fully-relativistic electronic structure calculation \cite{vir11} finds the required band-ordering, although in this particular case the $\Gamma_7$ bands and $\Gamma_6$ have switched places with respect to the normal ordering [c.f. Fig.~\ref{fig1}] and thus $\Delta_0 <0$. This reversed order originates from a small but
significant contribution of Hg $5d$ orbitals whose spin-orbit (SO) coupling dominates over the sulfur $3p$ states and reverses its sign \cite{del02} by an amount sufficient enough to create a small gap thus rendering $\beta$-HgS a stoichiometric strong 3D TI .

{\it Inversion invariant effective Hamiltonian --} For an analysis of the topological surface states one can rely on effective low-energy theories that are based upon a ${\bf k} \cdot {\bf p}$ expansion of the lowest energy bands around the $\Gamma$ point. In the simplest case, as for instance  Bi$_2$Se$_3$, the ${\bf k} \cdot {\bf p}$ theory yields a Dirac-like Hamiltonian \cite{zha09,liu10} with the Dirac-mass representing the energy gap and a change of its sign corresponding to a nontrivial level crossing. 
The same approach applies to the mercury chalcogenides and has correctly predicted the Quantum Spin Hall effect in HgTe/CdTe quantum wells \cite{ber06}. The generic form of a low-energy ${\bf k} \cdot {\bf p}$ expansion at the BZ center $\Gamma$ for semiconductors with a zinc-blende crystal structure is given by the Kane model Hamiltonian \cite{win03}. Previous work has focused predominantly on the bands responsible for the level crossing, the $\Gamma_6$ and $\Gamma_8$ bands \cite{dai08,chu11}. While such an analysis is capable of correctly describing the topological characteristics of HgTe and its consequences for surface excitations, here we consider the full eight-band Kane model Hamiltonian which takes into account the $\Gamma_6$, $\Gamma_7$ and $\Gamma_8$ bands and correctly describes the band ordering near the Brillouin zone centre of the series of mercury chalcogenides HgX once the spin-orbit splitting energy $\Delta_0$ is varied. 
This allows us to smoothly connect from the intrinsic, fully gapped, TI regime realised in $\beta$-HgS to the putative helical semimetal regime for $\Delta_0>0$ and analyse the fate of the resulting topological surface states. Even more $\Delta_0$  not only provides a convenient tuning parameter, its variation represents the physical effect of biaxial  strain fields. Simultaneous application of two stress fields directed along the $[100]$ ($[010]$) and $[001]$ directions will under specific conditions [see the Supplemental Material] preserve the degeneracy at the $\Gamma$ point among the LH and HH bands, but renormalize the SO energy $\Delta_0$. 

 \begin{figure}
 \includegraphics[width=\columnwidth]{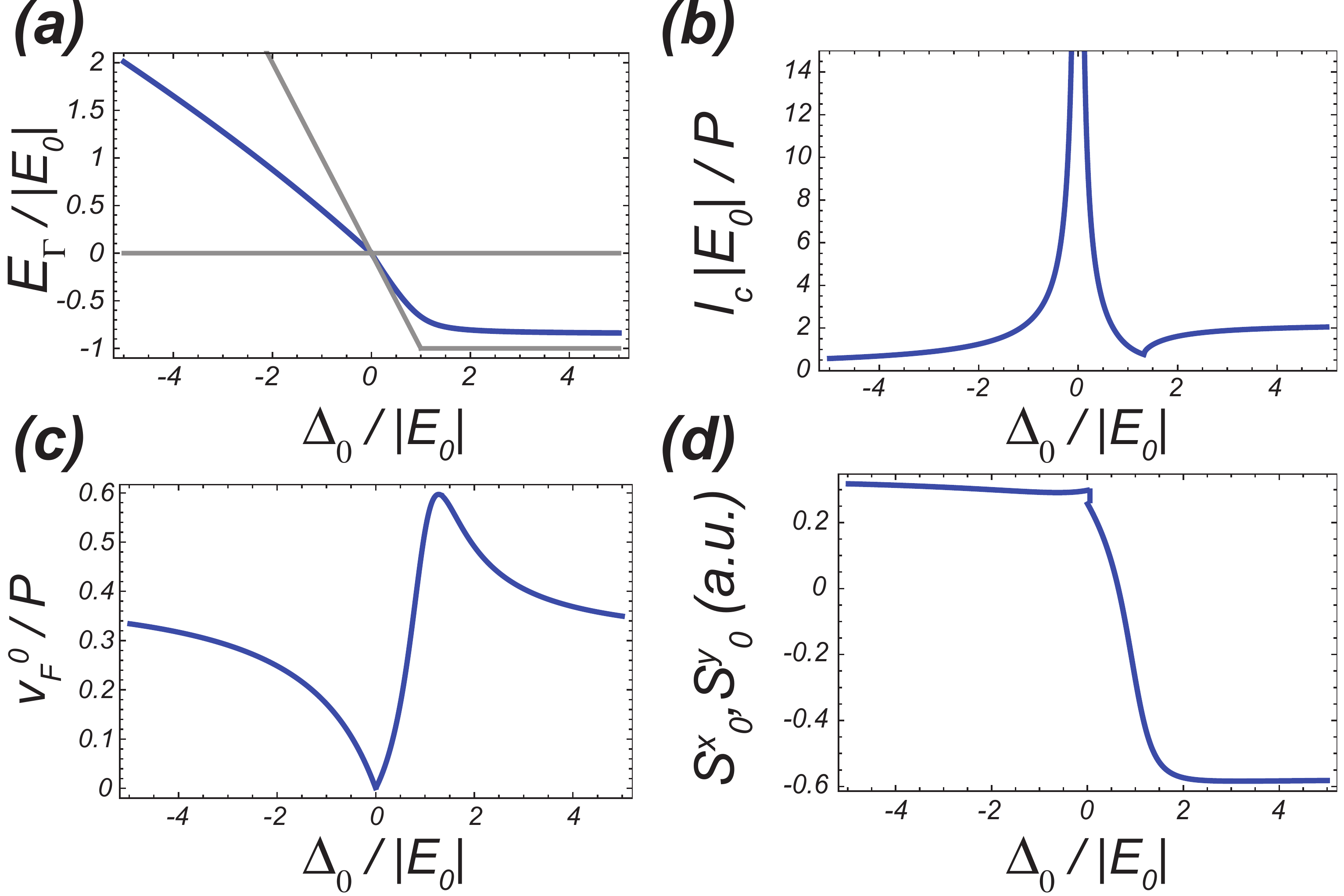}
 \caption{(a) Behavior of the energy of the surface Dirac point measured in units of the gap $|E_0|$ as a function of the ratio among the spin-orbit splitting $\Delta_0$ and $|E_0|$  in the absence of broken inversion symmetry. The gray lines correspond to the conduction and valence bulk band-edges by artificially removing the parasitic HH bands. (b) Behavior of the decay length for the surface states at the centre of the surface Brillouin zone as a function of $\Delta_0 / |E_0|$. (c) Same for the Fermi velocity of the surface Dirac cones $v_F^{0}$ measured in units of the band structure parameter $P$. (d) Behavior of the spin constants  $S_0^{x,y}$ as a function of the spin-splitting energy $\Delta_0$.}
 \label{fig2}
 \end{figure}

To establish the helical semimetal state in the cubic mercury chalcogenides  when inversion symmetry is preserved, we explicitly calculate the [001] surface states of the eight-band Kane model Hamiltonian by neglecting BIS effects on the half-space $z>0$ \cite{liu10} with open boundary conditions. For illustration of the physics we take for simplicity the HgTe band structure parameters at $T=0$ K \cite{nov05}.
At the $\Gamma$ point of the surface BZ, the Kane model Hamiltonian predicts the HH bands to be completely decoupled from the other bands. This decoupling guarantees 
the absence of any mixing among the parasitic HH bands and the topological surface states resulting from the TI bulk. 
The corresponding  part of the Hamiltonian is block diagonal with the two blocks for the chalcogen $p$-type (mercury $s$-type) states of total angular momentum $J_z = 1/2$ and $J_z = -1/2$  respectively. The eigenstates have therefore the form 
$\Psi^{\uparrow} (z)  = ( \psi_0^{\uparrow} , {\bf 0})^{T}$  and $\Psi^{\downarrow} (z)  = ({\bf 0},  \psi_0^{\downarrow} )^{T}$
where $\psi_0^{\uparrow,\downarrow}$ is a three-dimensional spinor and ${\bf 0}$ is a five component zero vector. For the surface states, the wavefunction $\psi_0^{\uparrow,\downarrow} (z)$ is localized at the [001] surface in which case $\Psi^{\uparrow,\downarrow}$ play the role of a spin one-half surface Kramer's doublet [see the Supplemental Material].

Fig.\ref{fig2}(a) shows the energy of the surface Kramer's doublet as a function of the ratio among the SO splitting energy $\Delta_0$ and the $\Gamma_6-\Gamma_8$ gap $-E_0$. 
In the intrinsic, fully gapped, TI regime, $\Delta_0 <0$, the surface state's energy resides in the direct bulk insulating gap at the $\Gamma$ point. In the $\Delta_0 < 0$ regime, instead, the surface Kramer's doublet energy lies below the zero energy HH band-edge but resides in the band-gap of the TI bulk realised by  the $\Gamma_{6,7}$-LH bands. 
Fig.\ref{fig2}(b) shows the behavior of the decay length of the surface states. 
We find that precisely at $\Delta_0 \equiv 0$ -- where the gap of the bulk TI closes --  the decay length diverges and thus the condition for the existence of the surface states is violated. For finite values of the spin-orbit splitting, instead, the existence condition for the surface states is fulfilled which is  guaranteed by the fact  that  a renormalizable surface state solution exists in the half-infinite space $z>0$. By projecting the bulk Hamiltonian onto the subspace of these two surface states \cite{zha09} , we obtain an effective surface Hamiltonian to the leading order of $k_{x,y}$ 
\begin{equation}
{\cal H}_{surf} (k_x, k_y) = E_\Gamma \, {\cal I} + v_F^{0}  \left( \sigma_x k_y - \sigma_y k_x \right), 
\end{equation} 
with the Fermi velocity $v_F^0$ whose behavior as a function of the spin-orbit splitting is shown in Fig.\ref{fig2}(c). 
That the $\sigma$ matrices in the effective surface model Hamiltonian  are proportional to the real spin can be shown by projecting the total angular momentum operators $J_{x,y,z}$ onto the surface state subspace. Independent of the spin-orbit splitting energy, we do find that $\bra{\Psi} J_{x,y,z} \ket{\Psi} \equiv S_0^{x,y,z} \sigma_{x,y,z} $ where $S_0^z \equiv 1/2$ whereas $S_0^{x} \equiv S_0^{y}$ with a finite value whose behavior as a function of $\Delta_0$ is shown in Fig.\ref{fig2}(d). As a result, the surface states show a linear dispersion with helical spin-textures left-handed for the surface conduction band and right-handed for the surface valence band proving the spin-momentum locking of the surface state solutions. 

  \begin{figure}
  \includegraphics[width=\columnwidth]{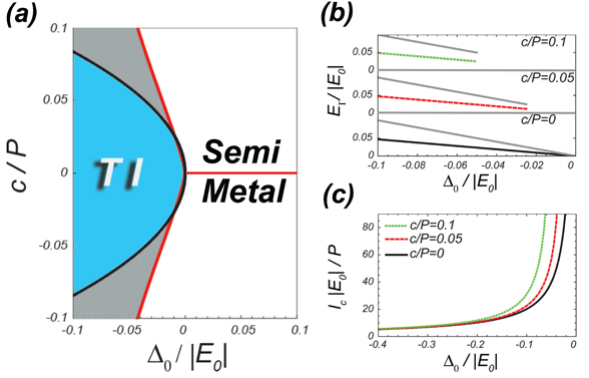}
  \caption{(a) Phase diagram for the existence of renormalizable surface state solutions in the [001] surface of HgX mercury chalcogenide compounds as obtained from the eight-band Kane model Hamiltonian. The gray area corresponds to regions with topological surface states at the BZ centre  where the opening of the indirect bulk band gap cannot be correctly captured by the Kane model . (b) Behavior of the energy of the surface Dirac point measured in units of the gap $|E_0|$ as a function of the ratio among the spin-orbit splitting $\Delta_0$ and $|E_0|$  for different strengths of the linear in ${\bf k}$ BIS terms. The Dirac point energy always lies in the gap at the $\Gamma$ point among the $\Gamma_8$ and the SO split-off bands $\Gamma_7$ whose band-edges are represent by the gray lines. (c) Same for the decay length of the surface states.
}
 \label{fig3}
 \end{figure} 

{\it Broken inversion symmetry --} 
Having established that in presence of inversion symmetry, the series of cubic mercury chalcogenides will either be in the strong 3D TI or in the helical semimetal state, we now  take into account the intrinsic BIS of the zinc-blende crystal structure.  From a ${\bf k.p}$ perspective, the BIS allows for additional terms in the bulk Hamiltonian once the point group symmetry is reduced to $D_{2d}$ \cite{dai08,win03}. In the valence band block ${\cal H}^{8 , 8}$ of the Kane model Hamiltonian [see the Supplemental Material] the BIS indeed yields an additional term \cite{dre55} ${\cal H}_{BIS}^{8 ,  8} = c \left[ \left\{ J_x , J_y^2 - J_z^2 \right\}  k_x + c.p. \right] / \sqrt{3} $ and a similar term in the ${\cal H}^{8 , 7}$ block ${\cal H}_{BIS}^{8 ,  7} =  - i \sqrt{3} c \left[  T_{y z}^{\dagger} k_x + c.p. \right]$. The presence of this linear in ${\bf k}$ additional terms stems from bilinear terms consisting of ${\bf k.p}$ and SO interaction with the uppermost $d$ core levels \cite{car86}. As a result, the parameter $c$ is an elementary parameter of the Kane model that unlike the higher order spin splitting terms induced by BIS cannot be expressed in terms of the extended Kane model \cite{win03}.  Because of the smallness of the elementary parameter $c \simeq $ 80  meV \AA \cite{car86} as compared to the linear parameter coupling $P \simeq$ 8 eV \AA $\,$for HgTe, the conventional wisdom \cite{nov05} is that the BIS effect is very small in mercury chalcogenides and can be therefore safely neglected. 

We do find, however, that the BIS  has drastic consequences on both the existence conditions  and the dispersion of the surface states. Independent of the actual $c$ value, indeed, the BIS-induced linear in ${\bf k}$ terms couple the HH with the TI bulk at the $\Gamma$ point of the surface BZ. As a result, the BIS leads to an effective hybridization among the topological surface states and the parasitic HH bands. One would then expect that  whenever the topological surface states overlap in momentum and energy with the parasitic HH bands they should be pushed away. And indeed we find that for positive values of the spin-orbit splitting $\Delta_0$, in which case the energy of the surface Kramer doublet lies below the zero-energy HH band-edge, 
localized surface state wavefunctions $ \Psi^{\uparrow,\downarrow}$ at the BZ centre {\it do not} exist. 
In the $\Delta_0 <0$ regime instead, the BIS-induced hybridization should not be effective at the $\Gamma$ point 
from which one would expect that the existence of the surface Kramer's doublet should not be hampered independent of the actual values of the $c$ parameter and the spin-orbit splitting. 
 On the contrary we find that the existence of topological surface states is intrinsically related to the strength of the linear in ${\bf k}$ BIS terms and leads to the phase diagram shown in  Fig.\ref{fig3}(a).
Remarkably for small values of the BIS parameter $c$, renormalizable surface states appear only whenever   
the spin-orbit splitting is negative by an amount sufficient to create a full indirect band gap. Thus, even in the absence of an overlap in momentum and energy with the parasitic HH bands, the topological surface states are prevented in the absence of a full bulk band-gap proving that the helical semimetal state is completely suppressed by the BIS.

\begin{figure}
  \includegraphics[width=\columnwidth]{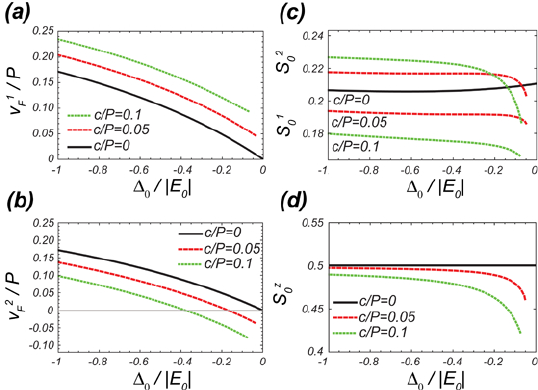}
  \caption{(a),(b) Behavior of the two non-equivalent Fermi velocities of the surface Dirac cones as a function of the spin-orbit splitting $\Delta_0$ for different values of the elementary parameter $c$ of the eight-band Kane model Hamiltonian. (c),(d) Same for the behavior of the spin constants $S_0^{1,2,z}$. }
  \label{fig4}
  \end{figure} 

Fig.\ref{fig3}(b) shows the behavior of the surface Kramer's doublet energy at the surface Brillouin zone centre for different values of the BIS elementary parameter $c$ 
when the intrinsic, fully gapped, TI regime is reached. It always lyes in  the bulk  gap at the zone centre  among the $\Gamma_8$ and the SO split-off $\Gamma_7$ bands. 
We also show [c.f. Fig.\ref{fig3}(c)]  the behaviour of the penetration depth of the surface states which, increasing the value of the spin-orbit splitting $\Delta_0$, increases 
and eventually diverges at the "topological phase transition" of the phase diagram in Fig.~\ref{fig3}(a).
 By projecting again the bulk Hamiltonian onto the subspace of the surface BZ surface states, we obtain that in the presence of BIS terms the effective surface Hamiltonian to the leading order of $k_{x,y}$ reads
\begin{equation}
{\cal H}_{surf} (k_x, k_y) = E_\Gamma \, {\cal I} + v_F^1 k_1 \sigma_1 - v_F^2 k_2 \sigma_2 , 
\end{equation} 
where $k_{1,2} = ( k_x \pm k_y  ) / \sqrt{2}$ and $\sigma_{1,2}$ are the corresponding rotated Pauli matrices $\sigma_{1,2} = ( \sigma_x \mp \sigma_y ) / \sqrt{2}$. As a result the surface Dirac cones are anisotropic ($v_F^1 \neq v_F^2$) along the diagonal directions of the surface BZ 
as can be shown by a two-dimensional ${\bf k.p}$ analysis [see the Supplemental Material] and 
 in perfect agreement with the density functional electronic structure calculations in $\beta$-HgS \cite{vir11}. In addition, the spin-momentum locking of the surface states is guaranteed by the fact that by projecting the $\pi/4$ rotated total angular momentum operators $J_{1,2,z}$ onto the subspace of the surface states at the BZ centre we find $\bra{\Psi} J_{1,2,z} \ket{\Psi} \equiv S_0^{1,2,z} \sigma_{1,2,z}$ with $S_0^{1,2,z}$ some constants the behaviour of which, as function of the spin-orbit splitting $\Delta_0$ is shown in Fig.\ref{fig4}(c),(d). 
Fig.\ref{fig4}(a),(b) show the behaviour of the two inequivalent Fermi velocities for different values of the BIS parameter $c$. It is evident that for $\Delta_0 \ll E_0$, the surface Dirac cone is strongly anisotropic with a large dispersion along the diagonal $k_1$ and a nearly flat band along the perpendicular direction. By varying the strength of the spin-orbit splitting energy, we find a critical value of the spin-orbit splitting $\Delta_0^c$ where the degree of anisotropy $v_F^1 / v_F^2$ diverges 
and the surface states will be completely one-dimensional.  
 Even more, the fact that only one of the two non-equivalent Fermi velocities changes sign, implies a change in the handedness of the surface Dirac cone -- left-handed for $\Delta_0 < \Delta_0^c $ and right-handed for  $\Delta_0 > \Delta_0^c$ in the surface conduction band. 
Therefore, a suitable application of anisotropic biaxial stresses  can induce a flip of chirality which would immediately manifest itself as a sign change of the quantized Hall conductance in the presence of a time-reversal symmetry breaking perturbation at the surface.

{\it Conclusions --} Coexistence of bulk metallic states with topological surface states can be encountered in a large class of materials. Inverted zero-gap semiconductors fall into this class and provide a prominent example of an helical semimetal. We have shown here that while such a topological state of matter can be established in crystals with inversion symmetry, a breaking of the bulk inversion symmetry initiates a bulk-surface state struggle in which the topological surface states, and thereby the helical semimetallic state as a whole, perish. In the intrinsic, fully gapped, TI regime, the broken inversion symmetry strongly renormalizes the anisotropy of the group velocity of the surface Dirac fermions similarly to graphene superlattices \cite{par08np,par08nl}. This might be a relevant feature for spin conduction experiments where a large anisotropy of the surface Dirac cone has been predicted to lead to very large spin lifetimes \cite{sac12}. 

The authors thank M. Richter and F. Virot for very fruitful  discussions.

\clearpage
\begin{appendix}
\section{Appendix A: Kanel model Hamiltonian}
\noindent

As long as the intrinsic bulk-inversion asymmetry of the zinc-blende crystal structure is not taken into account, the eight-band Kane model Hamiltonian reads
\begin{equation}
{\cal H}= \left( \begin{array}{ccc} {\cal H}^{6 , 6} & {\cal H}^{6 , 8} & {\cal H}^{6 , 7} \\ 
 \cdot & {\cal H}^{8 , 8} &  {\cal H}^{8 , 7} \\
 \cdot & \cdot  & {\cal H}^{ 7 , 7} 
\end{array}
\right)
\label{eq:hamiltonian}
\end{equation}
with  the expression of the Hamiltonian subblocks ${\cal H}^{\alpha , \beta}$ listed in Table \ref{tab:table1}. They are expressed in terms 
of the usual Pauli matrices $\sigma_{x,y,z}$ , the 
$J=3/2$ angular momentum matrices $J_x = \sqrt{3} / 2 \,  {\cal I} \otimes \sigma_x + ( \sigma_x \otimes \sigma_x + \sigma_y \otimes \sigma_y ) / 2$, $J_y = \sqrt{3} / 2 \,  {\cal I} \otimes \sigma_y  + ( \sigma_y \otimes \sigma_x - \sigma_x \otimes \sigma_y ) / 2$, $J_z = \sigma_z \otimes {\cal I} + {\cal I} \otimes \sigma_z  / 2$ with ${\cal I}$ the identity matrix and the following  $T_i$ matrices:
\begin{eqnarray*} 
T_x& = &\dfrac{1}{3 \sqrt{2}} \left( \begin{array}{cccc}- \sqrt{3} & 0 & 1 & 0 \\ 0 & -1 & 0 & \sqrt{3} \end{array} \right) \\
T_y &= & \dfrac{-i}{3 \sqrt{2}} \left( \begin{array}{cccc} \sqrt{3} & 0 & 1 & 0 \\ 0 & 1 & 0 & \sqrt{3} \end{array} \right) \\
T_z &= &\dfrac{\sqrt{2}}{3} \left( \begin{array}{cccc} 0 & 1 & 0 & 0 \\ 0 & 0 & 1 & 0 \end{array} \right) \\
T_{xx}&=& \dfrac{1}{3 \sqrt{2}} \left( \begin{array}{cccc} 0 & -1 & 0 & \sqrt{3} \\ -\sqrt{3} & 0 & 1 & 0 \end{array} \right) \\
T_{yy}&=& \dfrac{1}{3 \sqrt{2}} \left( \begin{array}{cccc} 0 & -1 & 0 & -\sqrt{3} \\ \sqrt{3} & 0 & 1 & 0 \end{array} \right) \\
T_{zz} & =& \dfrac{\sqrt{2}}{3} \left( \begin{array}{cccc} 0 & 1 & 0 & 0 \\ 0 & 0 & -1 & 0 \end{array} \right) \\
T_{yz} &=& \dfrac{i}{2 \sqrt{6}} \left( \begin{array}{cccc} -1 & 0 & -\sqrt{3} & 0 \\ 0 & \sqrt{3}  & 0 & 1 \end{array} \right) \\
T_{zx} &=&  \dfrac{1}{2 \sqrt{6}} \left( \begin{array}{cccc} -1 & 0 & \sqrt{3} & 0 \\ 0 & \sqrt{3}  & 0 & -1 \end{array} \right) \\
T_{xy} &=&  \dfrac{i}{ \sqrt{6}} \left( \begin{array}{cccc} 0 & 0 & 0 & -1 \\ -1 & 0  & 0 & 0 \end{array} \right) \\
\end{eqnarray*}

The parameters $F, \gamma_1, \gamma_2, \gamma_3$ describe the coupling to remote bands and are, as well as $P, E_0$, material-specific parameters. 
We have considered for simplicity the axial approximation $\overline{\gamma} = ( \gamma_2 + \gamma_3 ) / 2$ with the warping parameter $\mu= (\gamma_3-\gamma_2) / 2 \equiv 0$  in order to make the bulk band structure isotropic in the $k_{x,y}$ plane \cite{nov05}.

\begin{table}[tb]
\centering
 \caption{Expressions of the Kane model in the axial approximation. Here $\left\{A, B \right\}$ denotes the anticommutator for the $A,B$ operators,  $c.p.$ cyclic permutations of the preceding term and we defined $B= \hbar^2 / ( 2 m_0)$ with $m_0$ the free electron mass. We also list the band structure parameters for HgTe of Ref.\cite{nov05} .}
\begin{ruledtabular}
\begin{tabular}{ccc}
Hamiltonian blocks & & {\bf k $\cdot$ p} interactions \\
\hline
${\cal H}^{6 , 6}$ & & $E_0 + B (2F + 1 ) {\bf k}^2 $ \\ [1ex]
${\cal H}^{6 , 8}$ & & $\sqrt{3} P\,  {\bf T \, \cdot \, k}$ \\ [1ex]
${\cal H}^{6 , 7}$ & & $-\frac{1}{\sqrt{3}} \, P \, {\boldsymbol \sigma} {\bf \cdot} {\bf k} $ \\ [1ex] 
& &  $-B \gamma_1 {\bf k}^2 + 2 B \overline{\gamma} \left[ \left(J_x^2 - \frac{J^2}{3} \right) k_x^2 + c.p. \right] $ \\ [-1ex]
\raisebox{1.5ex}{$ {\cal H}^{8 , 8} $} & & $ + B \overline{\gamma} \left[ \left\{ J_x , J_y \right\} \left\{k_x , k_y \right\} + c.p. \right] $ \\ [1ex]
 & & $6 B \overline{\gamma} \left[ \left(T_{xx}^{\dagger} k_x^2 + c.p.  \right) + \right. $  \\   [-1ex]
\raisebox{1.5ex}{$ {\cal H}^{8 , 7}$} & & $\left. \left( T_{xy}^{\dagger} \left\{k_x , k_y \right\} + c.p. \right) \right]$ \\ [1ex]
${\cal H}^{7 , 7}$ & & $-\Delta_0 - B \gamma_{1} {\bf k}^2$ \\
\end{tabular}
\end{ruledtabular}
\vspace{.2cm}
\begin{ruledtabular}
\begin{tabular}{cccccc} 
$E_0$ & $F$ & $P^2/ B $ & $\gamma_1$ & $\gamma_2$ & $\gamma_3$ \\
\hline
-0.3 eV & 0 & 18.8 eV & 4.1 & 0.5 & 1.3 
\end{tabular}
\end{ruledtabular}
\label{tab:table1}
\end{table}

\begin{table}[tb]
\centering
 \caption{Strain-induced terms in the eight-band Kane model Hamiltonian}
\begin{ruledtabular}
\begin{tabular}{ccc}
Hamiltonian blocks & &strain-induced interactions \\
\hline
${\cal H}^{6 , 6}$ & & $C_{1} Tr \epsilon$ \\ [1ex]
& & $D_d Tr \epsilon + \frac{2}{3} D_u \left[ \left(J_x^2 - \frac{1}{3} J^2 \right) \epsilon_{xx} + c.p. \right]  $\\ [-1ex]
\raisebox{1.5ex}{$ {\cal H}^{8 , 8} $}  & & $\frac{2}{3} D_u^{\prime} \left[ \left\{J_x , J_y \right\} \epsilon_{x y} + c.p. \right]$  \\[1ex]
${\cal H}^{7 \, , 7}$ & & $D_d Tr \epsilon$  \\ [1ex] 
\end{tabular}
\end{ruledtabular}
\label{tab:table2}
\end{table}
Effects of strain can be taken into consideration by applying the formalism of Bir and Pikus. They lead to additional terms in the eight-band Kane model Hamiltonian  proportional to the strain tensor ${\boldsymbol \epsilon}$ and 
expressed in terms of the  valence band deformation potentials  $C_{1}, D_u, D_u^{\prime}$  [see Table\ref{tab:table2}]  
The strain-induced interactions then lead to a change in the band-edges with the degeneracy at the $\Gamma$ point among the LH and HH bands that is preserved for a strain field with $\epsilon_{xx} + \epsilon_{yy} = 2 \epsilon_{zz}$. 
This condition can be achieved, for instance by a simultaneous application of two uniaxial stresses along the [100] ([010]) and [001] direction of magnitudes $X$ ($Y$) and $Z$ respectively with $X,Y = 2 Z$ in which case $Tr \epsilon = 3 \left( S_{11} + 2 S_{12} \right) Z$ where $S_{11},S_{12}$ are the elastic compliance constants. 

\section{Appendix B: Surface state solutions}
We solve the eight-band Kane model Hamiltonian on the half-infinite space $z>0$ with open boundary conditions. We first consider our model Hamiltonian Eq.\ref{eq:hamiltonian} at the centre of the surface Brillouin zone ($k_{x,y} \equiv 0$)  where the degeneracy of the surface bands is protected by time-reversal symmetry. We therefore replace $k_z \rightarrow -i \partial_z$ and obtain the Schr\"odinger equation 
${\cal H}_{k_{x,y} \equiv 0} \left( k_z \rightarrow -i \partial_z \right) \Psi(z) \equiv E_{\Gamma} \Psi(z)$ 
In the absence of BIS, the eigenstates take the form
\begin{equation} 
\Psi^{\uparrow} (z) = \left( \begin{array}{c} \psi_0^{\uparrow} \\ {\bf 0} \end{array} \right) \hspace{1cm} \Psi^{\downarrow} (z) = \left( \begin{array}{c} {\bf 0} \\ \psi_0^{\downarrow}  \end{array} \right). 
\end{equation}
where $\psi_0^{\uparrow,\downarrow}$ is a three-dimensional spinor and ${\bf 0}$ is a five component zero vector. 
Obviously to obtain the surface states, the wavefunction $\psi_0^{\uparrow,\downarrow} (z)$ should be localised at the [001] surface. We therefore put a trial spinorial wavefunction of the form $\psi_0^{\uparrow,\downarrow} (z) = \psi_\lambda^{\uparrow,\downarrow} \,  \mathrm{e}^{\lambda z}$  into the Schr\"odinger equation thereby obtaining two eigenvalue equations with a unique secular equation the solution of which yields the general wavefunction solution 
$$\psi_0^{\uparrow,\downarrow}  (z) = \sum_{\alpha=1}^{3} \sum_{\beta = \pm} C_{\alpha \beta} \psi_{\alpha \beta}^{\uparrow, \downarrow} \, \mathrm{e}^{\beta \lambda_{\alpha} (E_{\Gamma}) \,  z}.  $$
The normalizability of the wavefunction in the $z>0$ region requires that the wavefunction contains only the three terms with $\beta$ negative and immediately yields the existence condition of the surface states ${\cal R} (\lambda_{\alpha}) \neq 0$ preventing the surface states to penetrate into the bulk and defines the decay length of the surface states $l_c = max\left\{ 1 /  {\cal R} (\lambda_\alpha) \right\} $.
Furthermore, applying the boundary condition $\psi_0^{\uparrow, \downarrow} (z =  0) \equiv 0$, we obtain a secular equation \cite{sha10} of the non-trivial solution for the coefficients $C_{\alpha \beta}$  that determines the energy of the two degenerate surface states at the surface Brillouin zone centre. 

When the intrinsic BIS of the zinc-blende crystal structure is taken into account, the eigenstates
of the Hamiltonian ${\cal H}_{k_{x,y} \equiv 0}$ 
have the form 
$\Psi^{\uparrow} (z)  = ( \psi_0^{\uparrow} , {\bf 0})^{T}$  and $\Psi^{\downarrow} (z)  = ({\bf 0},  \psi_0^{\downarrow} )^{T}$
with $\psi_0^{\uparrow,\downarrow}$ now a four-dimensional spinor. 
With this, the general solution can be written as $\psi_0^{\uparrow,\downarrow}  (z) = \sum_{\alpha=1}^{4} \sum_{\beta = \pm} C_{\alpha \beta} \psi_{\alpha \beta}^{\uparrow, \downarrow} \, \mathrm{e}^{\beta \lambda_{\alpha} (E_{\Gamma}) \,  z} $  with the condition of renormalizability of the wavefunction implying ${\cal R}(\lambda_\alpha) \neq 0$ and $\beta$ negative. 
For positive spin-splitting energy $\Delta_0$
the condition for the existence of the surface states is never verified since one of the $\lambda_{\alpha}$ is purely imaginary independent of the $c$ value. 
For negative spin-orbit splittings $\Delta_0$ instead,  a renormalizable surface state solution can exist provided a non-trivial solution for the coefficients $C_{\alpha \beta}$ can be found. 
 
\section{Appendix C: Two-dimensional ${\bf k.p}$ theory}
The point group of cubic mercury chalcogenides, is the group $T_d$, which does not contain the inversion operation. If one of these materials is terminated in the (001) direction, leading to a two-dimensional (001)-surface, the point group of that surface is reduced to ${\cal C}_{2v}$, consisting of a twofold rotation symmetry along the $z$ axis, and two mirror symmetries  $M_x: x\rightarrow -x $ and $M_y: y \rightarrow -y$  along the two diagonal $\Gamma-M$  directions of the surface BZ. By choosing as a natural basis for the Kramer doublet at the centre of the surface Brillouin zone,  the total angular momentum $J=\pm 1/2$, we have that the twofold rotational symmetry can be represented as $C_2= - i \sigma_z$.  Similarly we can represent the two mirror operations as $M_x= -i \sigma_x$ and $M_y=  - i \sigma_y$ by choosing the phases of $| \psi_{\uparrow,\downarrow} \rangle$  appropriately. Finally the anti unitary time-reversal operator is represented as usual as ${\cal T}= i \sigma_y {\cal K}$.
The Kramers doublet is split away from the centre of the surface BZ and the corresponding surface band structure can be studied  within the ${\bf k.p}$ framework. 
The form of the effective surface Hamiltonian  ${\cal H}({\bf k})$ is highly constrained by time-reversal and crystal symmetries. Indeed under $C_2$ and $M_{x,y}$, spin and momentum transform as follows: 
\begin{eqnarray} 
C_{2} &:& k_{x,y} \rightarrow -k_{x,y}, \sigma_{x,y} \rightarrow -\sigma_{x,y}, \sigma_z \rightarrow \sigma_z  \nonumber \\ 
M_x &:& k_x \rightarrow -k_x, k_y \rightarrow k_y, \sigma_x \rightarrow \sigma_x, \sigma_{y,z} \rightarrow -\sigma_{y,z} \nonumber  \\ 
M_y &=& M_x \,\, (x \rightarrow y )
\label{eq:sktransform} 
\end{eqnarray}
The Hamiltonian${\cal H}({\bf k})$ must be invariant under Eq.\ref{eq:sktransform}. In addition, time-reversal symmetry gives the constraint 
\begin{equation}
{\cal H}({\bf k})=\sigma_y {\cal H}^{\star} (-{\bf k}) \sigma_y 
\label{eq:trtransform} 
\end{equation}
As a result, we find that the effective Hamiltonian must take the following form up to second order in ${\bf k}$: 
\begin{equation}
{\cal H}({\bf k})_{2 v} =E_{2 v} ({\bf k}) \mathbb{I}+  v_F^{x} k_x   \sigma_y - v_F^{y} k_y  \sigma_x \label{eq:hamiltonianappendixc2v} 
 \end{equation}
 where $E_{2 v} ({\bf k})= \hbar^2 k_x^2 / (2 m_x)  +\hbar^2 k_y^2 / (2 m_y )$ generates particle-hole asymmetry. 
The Hamiltonian for the surface states at the (001) surface of a material with diamond crystal structure which instead possesses inversion symmetry can be easily calculated from Eq.\ref{eq:hamiltonianappendixc2v} by adding the additional constraint due to the four-fold rotational symmetry along the $z$ axis. As C$_4$ can be represented as $\mathrm{e}^{i \pi  \sigma_z / 4}$ we have that spin and momentum transform as 
\begin{equation}
C_4: k_x \rightarrow -k_y \,, k_y \rightarrow k_x \,, \sigma_x \rightarrow -\sigma_y \,, \sigma_y \rightarrow \sigma_x.
\end{equation}
As a result, the Hamiltonian for the corresponding bulk inversion symmetric material would read as 
\begin{equation}
{\cal H}({\bf k})_{4 v}= E_{4 v} ({\bf k}) \mathbb{I}+  v_F  \left(k_x  \sigma_y - k_y \sigma_x \right) 
 \end{equation}
where  $E_{4 v} ({\bf k})= \hbar^2 k^2 / (2 m) $. Therefore systems with inversion symmetry display the same emerging ${\cal U}$(1) rotational symmetry as is Bi$_2$Se$_3$ \cite{fu09}. 

 \end{appendix}


\begin{thebibliography}{10}

\bibitem{kan05}
C.~L. Kane and E.~J. Mele, Phys. Rev. Lett. {\bf 95},  226801  (2005).

\bibitem{kan05b}
C.~L. Kane and E.~J. Mele, Phys. Rev. Lett. {\bf 95},  146802  (2005).

\bibitem{ber06}
B.~A. Bernevig, T.~L. Hughes, and S.-C. Zhang, Science {\bf 314},  1757
  (2006).

\bibitem{wu06}
C. Wu, B.~A. Bernevig, and S.-C. Zhang, Phys. Rev. Lett. {\bf 96},  106401
  (2006).

\bibitem{kon07}
M. K\"onig, S. Wiedmann, C. Br?ne, A. Roth, H. Buhmann, L.~W. Molenkamp, X.-L.
  Qi, and S.-C. Zhang, Science {\bf 318},  766  (2007).

\bibitem{fu07}
L. Fu and C.~L. Kane, Phys. Rev. B {\bf 76},  045302  (2007).

\bibitem{fu07b}
L. Fu, C.~L. Kane, and E.~J. Mele, Phys. Rev. Lett. {\bf 98},  106803  (2007).

\bibitem{moo07}
J.~E. Moore and L. Balents, Phys. Rev. B {\bf 75},  121306  (2007).

\bibitem{zha09}
H. Zhang, C.-X. Liu, X.-L. Qi, X. Dai, Z. Fang, and S.-C. Zhang, Nat Phys {\bf
  5},  438  (2009).

\bibitem{hsi08}
D. Hsieh, D. Qian, L. Wray, Y. Xia, Y.~S. Hor, R.~J. Cava, and M.~Z. Hasan,
  Nature {\bf 452},  970  (2008).

\bibitem{xia09}
Y. Xia, D. Qian, D. Hsieh, L. Wray, A. Pal, H. Lin, A. Bansil, D. Grauer, Y.~S.
  Hor, R.~J. Cava, and M.~Z. Hasan, Nat Phys {\bf 5},  398  (2009).

\bibitem{che09}
Y.~L. Chen, J.~G. Analytis, J.-H. Chu, Z.~K. Liu, S.-K. Mo, X.~L. Qi, H.~J.
  Zhang, D.~H. Lu, X. Dai, Z. Fang, S.~C. Zhang, I.~R. Fisher, Z. Hussain, and
  Z.-X. Shen, Science {\bf 325},  178  (2009).

\bibitem{hsi09}
D. Hsieh, Y. Xia, L. Wray, D. Qian, A. Pal, J.~H. Dil, J. Osterwalder, F.
  Meier, G. Bihlmayer, C.~L. Kane, Y.~S. Hor, R.~J. Cava, and M.~Z. Hasan,
  Science {\bf 323},  919  (2009).

\bibitem{has10}
M.~Z. Hasan and C.~L. Kane, Rev. Mod. Phys. {\bf 82},  3045  (2010).

\bibitem{ras13}
B. Rasche, A. Isaeva, M. Ruck, S. Borisenko, V. Zabolotnyy, B. B{\"u}chner, K.
  Koepernik, C. Ortix, M. Richter, and J. van~den Brink, Nat Mater {\bf 12},
  422 (2013).

\bibitem{akh09}
A.~R. Akhmerov, J. Nilsson, and C.~W.~J. Beenakker, Phys. Rev. Lett. {\bf 102},
   216404  (2009).
   
\bibitem{eto10}
K. Eto, Z. Ren, A.~A. Taskin, K. Segawa, and Y. Ando, Phys. Rev. B {\bf 81}, 195309 (2010). 

\bibitem{ber10}
D.~L. Bergman and G. Refael, Phys. Rev. B {\bf 82},  195417  (2010).

\bibitem{bru11}
C. Br\"une, C.~X. Liu, E.~G. Novik, E.~M. Hankiewicz, H. Buhmann, Y.~L. Chen,
  X.~L. Qi, Z.~X. Shen, S.~C. Zhang, and L.~W. Molenkamp, Phys. Rev. Lett. {\bf
  106},  126803  (2011).

\bibitem{chu11}
R.-L. Chu, W.-Y. Shan, J. Lu, and S.-Q. Shen, Phys. Rev. B {\bf 83},  075110
  (2011).

\bibitem{par08np}
C.-H. Park, L. Yang, Y.-W. Son, M.~L. Cohen, and S.~G. Louie, Nat. Phys {\bf
  4},  213  (2008).

\bibitem{par08nl}
C.-H. Park, Y.-W. Son, L. Yang, M.~L. Cohen, and S.~G. Louie, Nano Letters {\bf
  8},  2920  (2008).

\bibitem{dai08}
X. Dai, T.~L. Hughes, X.-L. Qi, Z. Fang, and S.-C. Zhang, Phys. Rev. B {\bf
  77},  125319  (2008).

\bibitem{sva11}
A. Svane, N.~E. Christensen, M. Cardona, A.~N. Chantis, M. van Schilfgaarde,
  and T. Kotani, Phys. Rev. B {\bf 84},  205205  (2011).

\bibitem{vir11}
F. Virot, R. Hayn, M. Richter, and J. van~den Brink, Phys. Rev. Lett. {\bf
  106},  236806  (2011).

\bibitem{del02}
A. Delin, Phys. Rev. B {\bf 65},  153205  (2002).

\bibitem{liu10}
C.-X. Liu, X.-L. Qi, H. Zhang, X. Dai, Z. Fang, and S.-C. Zhang, Phys. Rev. B
  {\bf 82},  045122  (2010).

\bibitem{win03}
R. Winkler, {\em Spin-Orbit Coupling Effects in Two-Dimensional Electron and
  Hole Systems} (Springer, Berlin, 2003).

\bibitem{nov05}
E.~G. Novik, A. Pfeuffer-Jeschke, T. Jungwirth, V. Latussek, C.~R. Becker, G.
  Landwehr, H. Buhmann, and L.~W. Molenkamp, Phys. Rev. B {\bf 72},  035321
  (2005).

\bibitem{dre55}
G. Dresselhaus, Phys. Rev. {\bf 100},  580  (1955).

\bibitem{car86}
M. Cardona, N.~E. Christensen, and G. Fasol, Phys. Rev. Lett. {\bf 56},  2831
  (1986).

\bibitem{sac12}
V.~E. Sacksteder, S. Kettemann, Q. Wu, X. Dai, and Z. Fang, Phys. Rev. B {\bf
  85},  205303  (2012).

\bibitem{sha10}
W.-Y. Shan, H.-Z. Lu, and S.-Q. Shen, New Journal of Physics {\bf 12},  043048
  (2010).

\bibitem{fu09}
L. Fu, Phys. Rev. Lett. {\bf 103},  266801  (2009).

\end{thebibliography}
\end{document}